\documentclass[pdflatex,sn-mathphys-num]{sn-jnl}
\usepackage{float}%
\usepackage{graphicx}%
\usepackage{multirow}%
\usepackage{amsmath,amssymb,amsfonts}%
\usepackage{amsthm}%
\usepackage{mathrsfs}%
\usepackage[title]{appendix}%
\usepackage{xcolor}%
\usepackage{textcomp}%
\usepackage{manyfoot}%
\usepackage{booktabs}%
\usepackage{indentfirst}%
\usepackage{algorithm}%
\usepackage{algorithmicx}%
\usepackage{algpseudocode}%
\usepackage{listings}%
\usepackage{siunitx}%
\usepackage{makecell}%
\usepackage[utf8]{inputenc}
\theoremstyle{thmstyleone}%
\theoremstyle{thmstyletwo}%

\theoremstyle{thmstylethree}%

\begin{document}

\title[{Article Title}]{Physics-informed automated surface reconstructing via low-energy electron diffraction based on Bayesian optimization}

\author[1]{\fnm{Xiankang} \sur{Tang}}
\author[1]{\fnm{Ruiwen} \sur{Xie}}
\author[2]{\fnm{Jan P.} \sur{Hofmann}}
\author*[1]{\fnm{Hongbin} \sur{Zhang}}\email{hzhang@tmm.tu-darmstadt.de}

\affil[1]{\orgdiv{Theory of Magnetic Materials Group, Department of Materials and Geosciences}, \orgname{Technical University of Darmstadt}, \postcode{Otto-Berndt-Straße 3, 64287}, \city{Darmstadt}, \country{Germany}}
\affil[2]{\orgdiv{Surface Science Laboratory, Department of Materials and Geosciences}, \orgname{Technical University of Darmstadt}, \postcode{Peter-Grünberg-Straße 4, 64287}, \city{Darmstadt}, \country{Germany}}

\abstract{Low-energy electron diffraction (LEED) is a cornerstone technique for determining surface atomic structures~\cite{heldStructureDeterminationLowenergy2025}, yet the quantitative analysis of electron diffraction intensity as a function of incident electron energy---that is, LEED-\textit{I(V)} analysis---remains a complex inverse problem. In this work, we tackle quantitative LEED-\textit{I(V)} analysis based on physics-informed Bayesian optimization (BO). By embedding multiple scattering LEED forward models directly into a trust-region BO loop, our approach simultaneously optimizes both structural and experimental parameters, adaptively adjusting trust regions for efficient exploration of complex non-convex parameter spaces without manual intervention. The robustness and scalability of the approach are demonstrated using the Ag(100)-(1×1) and Fe\textsubscript{2}O\textsubscript{3}(1$\overline{1}$02)-(1×1) surfaces as examples. Our work establishes a general framework for solving inverse problems in various characterization techniques, unlocking a physics-informed efficient, reproducible, and autonomous paradigm.}

\keywords{low-energy electron diffraction-\textit{I(V)}, Bayesian optimization, surface reconstruction}

\maketitle

\section{Introduction}\label{sec1}

Surface reconstruction gives rise to a variety of emerging chemical and physical properties absent in bulk materials, profoundly influencing phenomena such as catalytic activity~\cite{zengSurfaceReconstructionRegulation2025, zhanSituCharacterizationMethod2025,zhangSituSurfaceReconstruction2023}, electronic structure~\cite{leeChargeDensityWave2023,yangDirectObservationStrong2024a,lion$v$Ga$_2$O$_3$001SurfaceReconstructions2025}, and magnetic ordering~\cite{spiridisFe3O4001FilmsFe0012006,stankaSurfaceReconstructionFe3O40012000}. 
Experimentally, the evolution of the surface structure during molecular beam epitaxy growth~\cite{shenDevelopmentSituCharacterization2024,khaireh-waliehDatadrivenAzimuthalRHEED2025a} is typically investigated using scanning tunneling microscopy (STM)~\cite{suchSTMNcAFMInvestigation2003}, atomic force microscopy~\cite{yamamotoDirectObservationSi11016x22020}, and diffraction-based methods~\cite{zhanSituCharacterizationMethod2025,moritzSurfaceStructureDetermination2022}. 
Among these, low-energy electron diffraction (LEED) plays an important role by providing direct and quantitative information on surface periodicity and atomic-scale structural reconstructions~\cite{setzerLEEDIntensitySurface1998,spiridisFe3O4001FilmsFe0012006,stankaSurfaceReconstructionFe3O40012000,heldStructureDeterminationLowenergy2025,moritzSurfaceStructureDetermination2022,chenGermaneneStructureEnhancement2021}. 
LEED produces characteristic intensity-energy ($I(V)$) curves that encode information about the atomic positions, and quantitative analysis of the LEED-$I(V)$ spectra can be used to determine the surface structure of specimens~\cite{hofmannAdsorptionChlorineRu0001A2012a,tianTwoDimensionalMetalPhosphorusNetwork2020,moritzSurfaceStructureDetermination2022,gavazaTheoryLowenergyElectron2007a,gavazaTheoryLowenergyElectron2007,moritzPerspectivesSurfaceStructure2009}.
Unlike other diffraction-based probes, such as surface X-ray diffraction (SXRD) or high-energy reflection electron diffraction, LEED’s advantage lies in its surface sensitivity, providing a clear and static overview of surface structure, and it can be easily performed in laboratory settings~\cite{hafezReviewGeometricInterpretation2022,moritzSurfaceStructureDetermination2022}. 

\begin{figure}[H]
\centering
\includegraphics[width=1\textwidth]{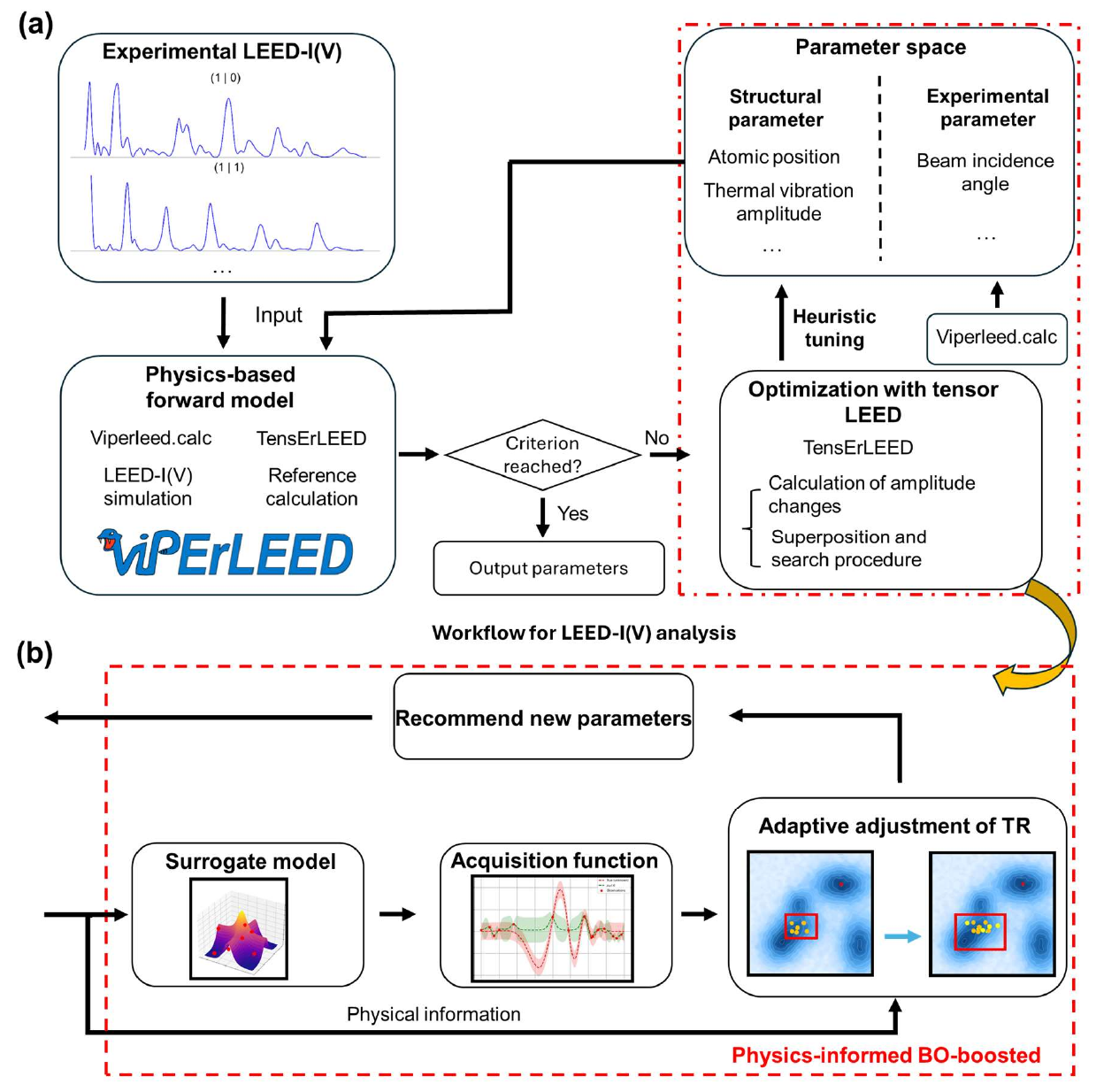}
\caption{\textbf{Workflow of ViperLEED.calc and proposed Bayesian optimization (BO) for LEED-$I(V)$ analysis}. (a) 
In ViperLEED.calc, the experimentally measured LEED-$I(V)$ curves are used as input. The software simulates corresponding LEED-$I(V)$ spectra based on structural and experimental parameters, evaluates the agreement using the \textit{R}-factor (Eq.~\ref{eq:min}), and generates new candidate structures through TensErLEED. The parameters are iteratively refined to minimize the \textit{R}-factor. In conventional practice, reliable results often require manual tuning of parameter ranges and optimization strategies based on human expertise. (b) In the BO framework, a surrogate model is employed to model the relationship between the parameters and the \textit{R}-factor, and an acquisition function guides the selection of the next candidate parameter set. The efficiency of parameter search is further enhanced by an adaptively adjusted trust region, thereby enabling efficient and adaptive exploration of the parameter space.}
\label{fig:workflow}
\end{figure}

However, the quantitative analysis of LEED-$I(V)$ represents an exceptionally challenging inverse problem. The experimentally measured spectra are linked to the surface structure through a highly nonlinear multiple scattering process~\cite{vanhoveSurfaceCrystallographyLEED1979,PendryLowEnergy}. Although the forward problem, {\it i.e.}, calculating LEED-$I(V)$ curves for a given structural model, has been well addressed using computational packages such as AQuaLEED~\cite{lachnittAQuaLEEDQuantitativeLEED}, EasyLEED~\cite{tiffeau-mayerAndimEasyleed2025}, CLEED~\cite{EmpascientificitCleedpy2025,heldRealisticMolecularDistortions1996}, and ViperLEED~\cite{kraushoferViPErLEEDPackageCalculation2024,schmidViPErLEEDPackageII2025}, the inverse problem of reconstructing surface structures from experimental data remains inherently complex and challenging. To tackle such challenges, researchers have developed the tensor LEED algorithm~\cite{rousTensorLEEDTechnique,rousTheoryTensorLEED1989}, which substantially enhances computational efficiency by applying perturbation theory to the diffraction matrix. For instance, ViperLEED.calc~\cite{kraushoferViPErLEEDPackageCalculation2024,schmidViPErLEEDPackageII2025}, integrating TensErLEED~\cite{blumFastLEEDIntensity2001}, provides a representative and widely adopted workflow for quantitative LEED-$I(V)$ analysis and surface structure determination, as illustrated in Fig. \ref{fig:workflow}a. Within this framework, a parameter space, encompassing surface atomic positions, atomic thermal vibration amplitudes (VIBROCC), electron beam incidence angle, among others, is explored by sequential simulation of LEED-$I(V)$ spectra, evaluating the \textit{R}-factor against experimental data, and generating new structures through perturbative updates. While successful~\cite{kraushoferViPErLEEDPackageCalculation2024,schmidViPErLEEDPackageII2025}, this method remains highly reliant on expert intervention: parameter ranges, step sizes, and optimization strategies must be manually specified, and structural and experimental parameters are typically optimized in a decoupled, sequential manner. Consequently, this heavy reliance on human expertise limits scalability for complex reconstructions and poses challenges for the reproducibility and robustness of the analysis.

Beyond perturbation-based frameworks, various non-gradient global optimization methods have been applied to LEED-$I(V)$ analysis, including genetic algorithms~\cite{dollGlobalOptimizationLEED1996}, fast simulated annealing~\cite{nascimentoFastSimulatedAnnealing2001,correiaGeneralizedSimulatedAnnealing2004}, differential evolution algorithms~\cite{nascimentoDifferentialEvolutionGlobal2015}, and parameter tree approaches~\cite{imreStructuralOptimizationTensor2025} that further enhance TensErLEED. Although these methods improve searching efficiency, they still remain highly dependent on human expertise. Recently, autoencoder-based methods~\cite{ivanovAutoencoderLatentSpace2024} have provided data-driven alternatives for analysis. However, their applicability is constrained by the requirement for large-scale, high-quality training datasets, limited generalization capability across experimental conditions, and an inherent focus on statistical correlations rather than the underlying scattering physics, which compromises the physical interpretability and reliability of the models outside the training domain. 

These limitations collectively motivate a fundamental new perspective: quantitative LEED-$I(V)$ analysis should be formulated as a physics-informed inverse problem, in which the full multiple scattering forward model is retained and embedded within a probabilistic inference framework. Incorporating physical simulations into machine learning models offers a natural pathway for the optimization process, thereby reducing data requirements, enhancing generalizability, and ensuring the validity of physical solutions~\cite{mengWhenPhysicsMeets2025,wangWhenPhysicsinformedData2023}. From this perspective, Bayesian optimization (BO) provides a particularly suitable framework, as it incorporates uncertainty quantification mechanisms and efficiently navigates the strongly coupled, high-dimensional parameter space~\cite{zhangAutonomousAtomicHamiltonian2023a,abuawwadKalmanFilterEnhanced2025, iwamitsuReplicaExchangeMonteCarlo2021, shinotsukaSampleStructurePrediction2023, machidaBayesianEstimationXPS2021,maEmittanceMinimizationAberration2025a,maEmittanceMinimizationAberration2025}. Unlike expert-based heuristics, BO explicitly models epistemic uncertainty in sparsely sampled regions of the parameter space. A detailed description of the proposed BO method will be presented in the Method~\ref{sec2} section.

In this study, we implement a physics-informed BO framework to perform quantitative LEED‑$I(V)$ analysis and apply it to prototypical Ag(100)-(1×1) and Fe2O3\textsubscript{2}\textsubscript{3}(1$\overline{1}$02)-(1×1) surfaces, demonstrating the robustness of our algorithm in both cases representing simple and complex multi-parameter scenarios, respectively. Furthermore, we conducted explicit density functional theory (DFT) calculations to evaluate the structural energetics, further confirming the reliability of our algorithm. Therefore, our work provides a new perspective for quantitative LEED‑$I(V)$ analysis and paves the way for more efficient and automated experimental characterization. It is expected that our algorithm can also be extended for solving the other inverse problems in characterization techniques.

\section{Method}\label{sec2}

Quantitative LEED-$I(V)$ analysis can be formulated as a physics-constrained inverse problem, in which a set of parameters $\Theta$ needs to be retrieved. This can, in turn, be transformed into an optimization problem, namely, minimizing the \textit{R}-factor~\cite{imreStructuralOptimizationTensor2025}. In this work, we have selected the Pendry \textit{R}-factor~\cite{pendryReliabilityFactorsLEED1980,heinzElectronBasedMethods2013}:

\begin{equation}
    R_{\mathrm{P}}(\Theta) = \frac{\sum_{\mathbf{i}} \int \left( Y_{\mathbf{i}}^{\mathrm{th}}(E) - Y_{\mathbf{i}}^{\mathrm{expt}}(E) \right)^2 dE}{\sum_{\mathbf{i}} \int \left( \left(Y_{\mathbf{i}}^{\mathrm{th}}(E)\right)^2 + \left(Y_{\mathbf{i}}^{\mathrm{expt}}(E)\right)^2 \right) dE}
\label{eq:min}
\end{equation}
where $\mathbf{i}$ indexes the beams for which the \textit{R}-factor is calculated, and \(Y(E)\) is the Pendry \(Y\) function. \(Y(E)\) can be computed from the beam intensities \(I(E)\), their derivatives \(I^{\prime}(E)=\frac{dI}{dE}\), and the imaginary part of the inner potential \(V_{0i}\) as:

\begin{equation}
    Y(E)=\frac{I(E)/I^{\prime}(E)}{\left[I(E)/I^{\prime}(E)\right]^{2}+V_{0i}^{2}} 
\label{eq:2}
\end{equation}
Due to the highly nonlinear nature of multiple scattering and the coupling between different parameters, the resulting R$_\mathrm{P}$($\Theta$) exhibits a severely non-convex function in parameter space. 

In this work, the inverse problem is solved using a physics-informed approach, where the full multiple scattering simulation is retained throughout the entire optimization process. Corresponding $I(V)$ spectrum calculations are performed using ViperLEED~\cite{kraushoferViPErLEEDPackageCalculation2024,schmidViPErLEEDPackageII2025} and TensErLEED~\cite{rousTensorLEEDTechnique,rousTheoryTensorLEED1989} packages. Importantly, this study imposes no prior assumptions regarding the relative importance of different parameters. All parameters are treated equally within a unified search space, bypassing manually defined hierarchical optimization or additional heuristic tuning. Within the optimization loop, the simulation model acts as a black-box evaluator: beyond the intrinsic physical calculations of the tensor LEED formalism, no assumptions about gradients or linearization of the \textit{R}-factor landscape are made. This approach maintains full physical fidelity of the scattering process while enabling flexible integration with probabilistic optimization strategies.

To efficiently explore the high-dimensional parameter space under the constraint of expensive forward evaluations, we employ BO as a probabilistic search strategy. BO is an optimization algorithm designed for black-box objective functions that are computationally (or physically) expensive to evaluate~\cite{frazierTutorialBayesianOptimization2018}. This method uses a surrogate model, typically a Gaussian process (GP), to model the relationship between input parameters and output function values, while quantifying its uncertainty, thereby enabling modeling of objective functions from noisy data~\cite{frazierTutorialBayesianOptimization2018,rasmussenGaussianProcessesMachine2006}. The employment of appropriate acquisition functions allows BO to facilitate efficient sampling in the parameter space~\cite{frazierTutorialBayesianOptimization2018}. BO has been applied in materials science, including material inverse design, autonomous materials discovery, and automated analysis of experimental data (\textit {e.g.}, \cite{xueAcceleratedSearchMaterials2016a,yuanAcceleratedDiscoveryLarge2018a,balachandranExperimentalSearchHightemperature2018a,raoMachineLearningEnabled2022a,liSequentialClosedloopBayesian2024a,kusneOntheflyClosedloopMaterials2020a,gayon-lombardoDeepKernelBayesian2025}). Particularly in the field of materials characterization, BO has been widely implemented for the identification and analysis of spectroscopic~\cite{zhangAutonomousAtomicHamiltonian2023a,abuawwadKalmanFilterEnhanced2025,ozakiAutomatedCrystalStructure2020a,ranjanKvXrayEmission2025,iwamitsuReplicaExchangeMonteCarlo2021,shinotsukaSampleStructurePrediction2023,machidaBayesianEstimationXPS2021} or microscopy images\cite{pattisonBEACONAutomatedAberration2025b,kalininMachineLearningScanning2022b,maEmittanceMinimizationAberration2025a,maEmittanceMinimizationAberration2025}, and has been extended to automated experiments\cite{noackMethodsApplicationsAutonomous2023,szymanskiAutonomousLaboratoryAccelerated2023b,kalininMachineLearningAutomated2023a}.

The workflow of our physics-informed BO boosted autonomous LEED-$I(V)$ analysis is illustrated in Fig.~\ref{fig:workflow}b. Given the highly non-convex nature of the objective function, we adopt an adaptive trust-region (TR) strategy~\cite{erikssonScalableGlobalOptimization2019}, which dynamically adjusts the sampling range within the parameter space based on the observed improvement of the \textit{R}-factor, enabling robust navigation across local minima while maintaining numerical stability in high-dimensional spaces. As shown in Fig~\ref{fig:workflow}(b), the input parameter vector $\Theta$ and its corresponding \textit{R}-factor are used to train a surrogate model, specifically a SingleTaskGP~\cite{balandatBoTorchFrameworkEfficient2020}. The covariance matrix is constructed using the Matérn 2.5 kernel~\cite{gardnerGPyTorchBlackboxMatrixMatrix2021} implemented in Botorch~\cite{balandatBoTorchFrameworkEfficient2020} package. This GP model incorporates a relatively strong prior distribution on the kernel hyperparameters---\textit{i.e.}, assumptions about their likely ranges---to mitigate overfitting and enhance generalization performance and numerical stability when the covariates are normalized to the unit cube and outcomes are standardized (\textit{i.e.}, zero mean, unit variance)~\cite{rasmussenGaussianProcessesMachine2006,balandatBoTorchFrameworkEfficient2020}.

After training the GP, the acquisition function samples new candidate points within the TR in the current parameter space. In low-dimensional cases (\textit{i.e.}, the number of parameters fewer than 20), we recommend using parallel Upper Confidence Bound (qUCB)\cite{wilsonReparameterizationTrickAcquisition2017}, while in high-dimensional scenarios, we suggest combining Thompson sampling (TS) with multiple TRs to achieve sufficient sampling. The TR acts as an adaptive constraint on the search domain within the parameter space: when a newly evaluated batch of samples significantly reduces the \textit{R}-factor, the region expands to encourage broader exploration; conversely, if the improvement is insufficient, the region contracts to focus the search on more promising neighborhoods. Through this mechanism, the acquisition function is optimized over a controllable and well-behaved subset of the parameter space, thereby maintaining stable search performance even in high-dimensional problems.

This study selects two representative cases, the Ag(100)-(1×1) surface and the Fe\textsubscript{2}O\textsubscript{3}(1$\overline{1}$02)-(1×1) surface, to validate the effectiveness of the proposed method. The LEED-$I(V)$ data for both case studies are taken from the publicly accessible ViperLEED database~\cite{ExamplesViPErLEED0141}, and all crystal structure images presented in this study are generated using the VESTA software~\cite{mommaVESTA3Threedimensional2011a}. In the Ag(100)-(1×1) case, its low dimensionality allows for the use of a single TR to reduce computational overhead and avoid the complexity of maintaining multiple GPs. While a single region increases the risk of TuRBO getting trapped in a local optimum, the framework adopts a simple yet effective warm-restart strategy to reach a sufficiently converged minimum~\textit{R}-factor~\cite{martiMultistartMethodsCombinatorial2013,poloczekWarmStartingBayesian2016}. We saved all simulation results from each optimization run; once the TR shrinks to the predefined minimum size, the optimization is restarted from a newly sampled set of initial points generated under different random seeds, while retaining all previously simulated points in the training set. In practice, as the number of restarts increases, the probability of finding the global optimum also increases~\cite{leNonsmoothNonconvexStochastic2024}. In cases where the number of parameters is very large, and the parameter space is highly complex, \textit{i.e.}, Fe\textsubscript{2}O\textsubscript{3}(1$\overline{1}$02)-(1×1) surface, we employed multiple TRs to improve the efficiency of sampling. We further validated the physical validity of the algorithm-optimized structure using first-principles calculations. Single-point energy calculations of the structure were performed using Vienna \textit{ab initio} Simulation Package (VASP), with specific computational details provided in the Table~S1.

\section{Results and Discussion}\label{sec3}

\subsection{BO boosted autonomous structure determination  of Ag(100)-(1×1) surface}\label{subsec1}
\subsubsection{Autonomous structure reconstruction}\label{subsubsec1}

\begin{figure}[H]
\centering
\includegraphics[width=1\textwidth]{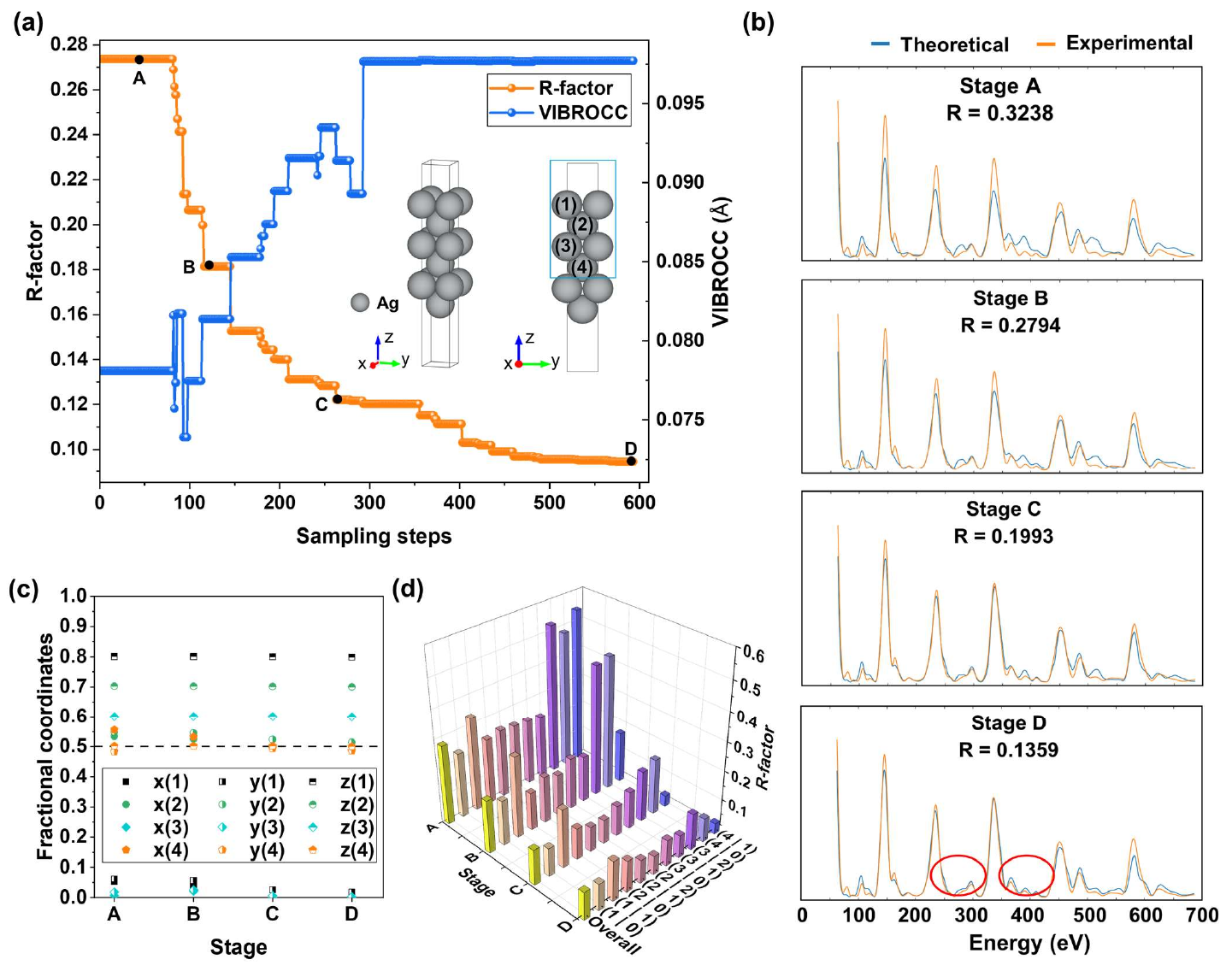}
\caption{\textbf{BO for LEED-$I(V)$ analysis of Ag(100)-(1×1) surface}. (a) Schematic of the Ag(100)-(1×1) surface structure. The blue box highlights the surface atoms (\textit{i.e.}, the atoms labeled (1) to (4)) whose positions are optimized during the optimization process. Also shown is the evolution of the \textit{R}-factor and VIBROCC (atomic thermal vibration amplitudes) as a function of the sampling steps. Note that the same horizontal axis scale is applied in Fig.~\ref{fig:energy}a and Fig.~\ref{fig:fe}. (b) LEED-$I(V)$ matching results for the $(1 \mid 1)$ diffraction beam at the four selected stages. In Stage D, the two red ellipses highlight representative detailed features in the LEED-$I(V)$. (c) Evolution of the fractional coordinates of the four atoms at the four selected stages, as marked in panel (a). (d) Variation of the overall \textit{R}-factor and the \textit{R}-factors of individual diffraction beams at the four selected stages, as marked in panel (a).}
\label{fig:rf}
\end{figure}

We consider Ag(100)-(1×1) surface structure determination as a proof-of-concept inverse problem to be solved using BO. Insets in Fig.~\ref{fig:rf}(a) display the reference surface structure, where the structural optimization is restricted to Ag atoms with fractional coordinates greater than 0.45 along the surface normal ($\textbf{\textit{z}}$-direction), corresponding to the atoms enclosed by the blue box in the $\textbf{\textit{y}}$-$\textbf{\textit{z}}$ view. The searching interval of 0.3~\si{\angstrom} is set for each atomic coordinate to be optimized, a deliberately expanded yet physically reasonable range that provides sufficient structural flexibility near the initial configuration while maintaining physical realism~\cite{kraushoferViPErLEEDPackageCalculation2024}. Additionally, the VIBROCC of all surface Ag atoms are simultaneously optimized within a variation range of 0.1~\si{\angstrom}. Overall, there are 13 parameters, constituting a representative test for automated LEED-$I(V)$ inverse quantification.

Despite the coupled nature of atomic coordinates and VIBROCC, the proposed framework autonomously converges to a physically meaningful low–\textit{R}-factor solution through fully adaptive TR control driven solely by \textit{R}-factor feedback. Figure~\ref{fig:rf}a displays the optimization trajectory, where both the \textit{R}-factor and VIBROCC are presented as functions of the sampling steps. In the initial stage of optimization, the algorithm explores within a large TR, starting from a relatively high \textit{R}-factor value of 0.2737. As the number of steps ({\it i.e.}, number of sampled configurations in the parameter space) increases, the \textit{R}-factor drops rapidly. Correspondingly, the TR adaptively shrinks, leading the algorithm to favor more refined sampling within the parameter space. Consequently, the decrease in the \textit{R}-factor gradually slows down and finally converges to 0.0946, comparable to results obtained by the ViPErLEED.calc~\cite{ExamplesViPErLEED0141} package after multi‑step optimization runs with manually adjusted hyperparameters. During this process, the evolution of the VIBROCC for all surface Ag atoms also exhibits a trend consistent with the R‑factor, initially rising rapidly before gradually converging. This indicates that the reduction in the \textit{R}-factor is not merely due to structural relaxation, thereby confirming the algorithm's effective exploration of the complex parameter space where VIBROCC and coordinates are coupled. It is noteworthy that this highly efficient exploration behavior is achieved entirely through TR adjustment driven by \textit{R}-factor feedback, without any manual intervention or predefined optimization hierarchy. This means that excellent LEED-$I(V)$ quantification can be achieved through a single automated task. In contrast, current state-of-the-art algorithms typically require at least two parameter adjustments, thereby reducing overall efficiency.

Representative configurations extracted along the optimization trajectory further illustrate the physically consistent convergence behavior. In Fig.~\ref{fig:rf}a, four representative configurations along the optimization trajectory are highlighted, with their corresponding spectral fitting results and atomic coordinate changes shown in Fig.~\ref{fig:rf}b and c, respectively. It is clear that during the optimization process, adjustments in atomic displacements primarily occur within the $\textbf{\textit{x}}$-$\textbf{\textit{y}}$ plane, while the positions of atoms along the $\textbf{\textit{z}}$-direction remain almost unchanged, as shown in Fig.~\ref{fig:rf}c. Given that LEED-$I(V)$ is highly sensitive to interlayer distances, this indicates that the initial model's layer spacing is already very close to the experimentally observed Bragg peak positions, implying that the actual energy-minimized configuration of the surface is primarily achieved through lateral atomic displacements rather than compression or stretching between layers. In the initial stage (from stage A to stage B), the algorithm samples within a large TR, which is equivalent to the initial coarse-grained screening performed by ViPErLEED.calc~\cite{ExamplesViPErLEED0141}. Additionally, the atoms shift towards the vertices (coordinate (0, 0)) and the center (coordinate (0.5, 0.5)) of the supercell on the horizontal plane (as shown in Fig.~\ref{fig:rf}c), contributing to a rapid decrease in the \textit{R}-factor, which is directly associated with the improved fitting of the relatively intense regions in the spectral curves, as shown in Fig.~\ref{fig:rf}b. In the intermediate stage (from stage B to stage C), the atomic coordinates are already close to their final positions. The algorithm starts to balance exploration and exploitation, performing adjustments within the now-reduced trust region, guiding the overall structure toward a stable configuration while progressively refining the spectral fit, as shown in Fig.~\ref{fig:rf}b, c. By the final stage (from stage C to stage D), the algorithm focuses on sampling within a highly localized parameter space, making only subtle atomic adjustments as shown in Fig.~
\ref{fig:rf}c, and concentrating on the precise matching of detailed features of LEED-$I(V)$ spectra, as marked in the last spectra matching in Fig.~\ref{fig:rf}b. This indicates that the algorithm has stably converged to a deep optimum in the parameter space. Improvements in matching for other beams are provided in Figs. S1–S4. The entire optimization process, including the evolution of the \textit{R}-factor, spectral fitting, and structural changes, can be clearly observed in the supplementary animation. Collectively, these results demonstrate that the algorithm autonomously achieves a process spanning from global exploration to local refinement, without relying on predefined optimization stages or expert intervention. Consequently, a single automated task is sufficient to achieve a solution quality that typically requires multiple manually adjusted optimization steps in conventional LEED-$I(V)$ analysis workflows.

During the optimization process, higher-order diffraction beams such as $(4 \mid 0)$ and $(4 \mid 1)$ exhibit a more significant systematic reduction in the \textit{R}-factor, as shown in Fig.~\ref{fig:rf}d. This behavior indicates that these reflections are inherently more responsive to structural refinements compared to lower-order beams like $(1 \mid 1)$. In particular, the $(4 \mid 1)$ diffraction beam experienced the largest reduction, with its \textit{R}-factor decreasing by more than half from stage A to stage B, indicating that it is the most sensitive to structural changes among all diffraction beams. Theoretically, the Pendry \textit{R}-factor (as defined in Eq:~\ref {eq:min}) is highly sensitive to structural changes~\cite{matererSurfaceStructuresLow1995}, and this sensitivity is even more pronounced for higher-order diffraction beams. Within the framework of LEED theory, this can be understood by considering the dynamical nature of multiple scattering: the intensity of each diffraction beam arises from the complex interference of multiple scattering paths. Higher-order diffraction beams correspond to larger scattering vectors, and their phase factors vary more rapidly~\cite{constantinouSensitivityConvergentBeam2019a}. Consequently, even small variations in atomic positions can induce amplified changes in their intensities. This makes these diffraction beams more sensitive to subtle structural optimizations and plays a crucial role in driving the reduction of the \textit{R}-factor during the optimization process.

Building on this autonomous convergence, a warm-starting strategy (see \ref{sec2} Method section) that reuses information accumulated from previous runs further improves the solution quality without altering the parameter search space. This strategy ultimately reduces the minimum \textit{R}-factor to 0.0772. The final optimized parameters and model hyperparameters are summarized in Table~\ref{tab:comparison} and Table S2, respectively. Notably, the algorithm does not depend on a fixed random seed for initialization. Instead, previously acquired information is continuously combined with newly generated random starting configurations, allowing the search process to leverage accumulated knowledge while preserving exploratory diversity. This design enhances robustness against metastable trapping and increases the probability of approaching the theoretical global optimum, consistent with the convergence characteristics discussed in the Methods~\cite{leNonsmoothNonconvexStochastic2024}.

\subsubsection{Energetic analysis of reconstructed structures}\label{subsubsec2}

\begin{figure}[H]
\centering
\includegraphics[width=1\textwidth]{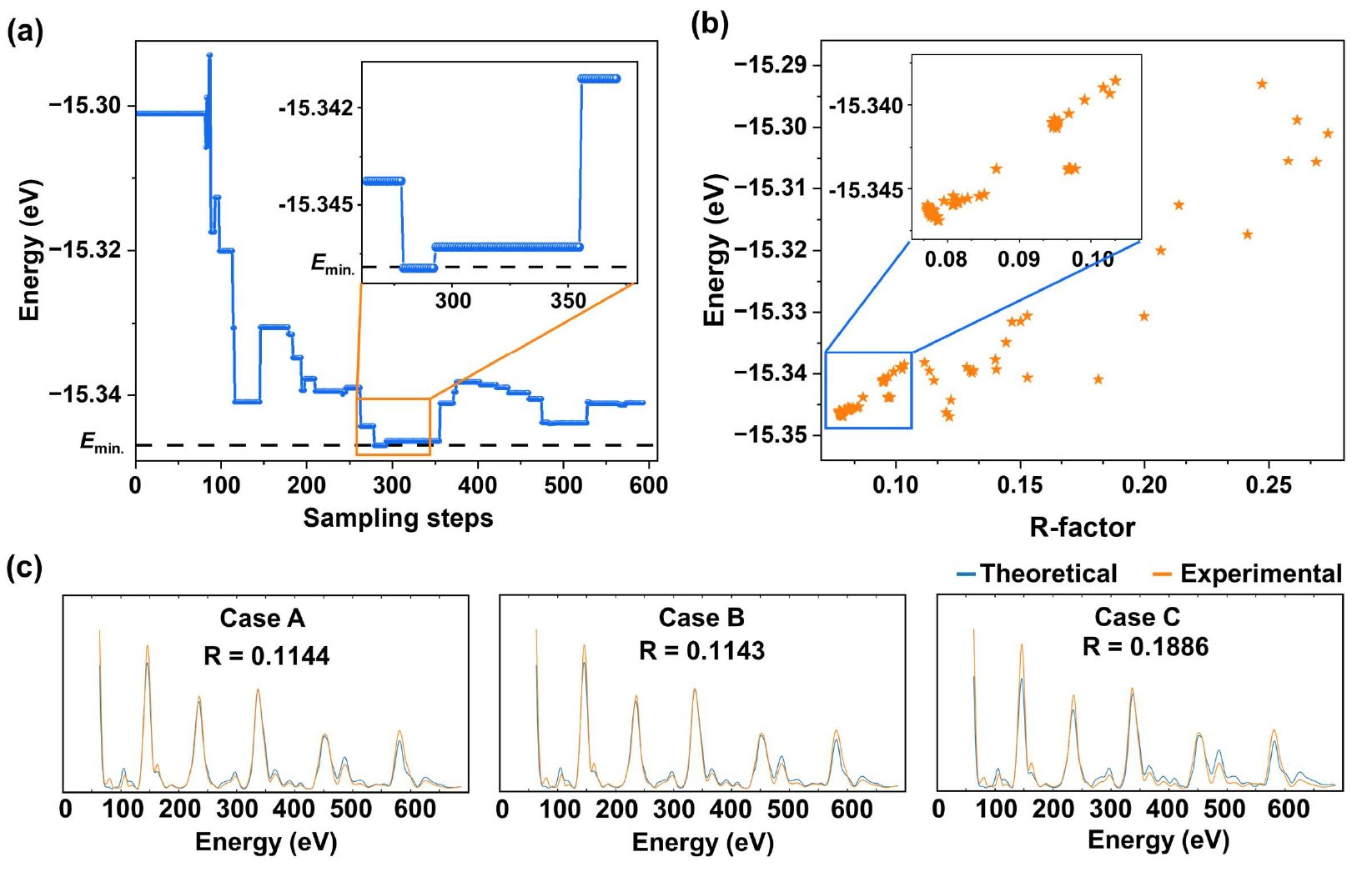}
\caption{\textbf{Energetic consistency and experimental validation of Bayesian-optimized Ag(100)-(1$\times$1) surface structures.}. (a) Evolution of the DFT total energy along the optimization trajectory, with an inset showing the lowest energy levels. (b) Correlation between the \textit{R}-factor and the DFT total energy for all sampled structures, showing that lower \textit{R}-factors are basically associated with energetically more stable configurations. (c) Comparison between experimental and theoretical LEED $I(V)$ curves for the representative $(1\bar{1}1)$ beam for three selected cases (A–C).  Case A and Case B are the results with the lowest energy and the lowest \textit{R}-factor, respectively, and details are listed in Table~\ref{tab:comparison}. Case C uses Case B's atomic structure but with VIBROCC (atomic thermal vibration amplitudes) fixed to bulk values, resulting in a significantly poorer agreement with experiment.}
\label{fig:energy}
\end{figure}

To validate the physical soundness of the quantification process, we performed DFT calculations to evaluate the energies of the sampled configurations, as illustrated in Fig.~\ref{fig:energy}a. Throughout the optimization process, the algorithm preserves physical consistency. Specifically, the total energies of the sampled configurations are overall decreasing with respect to the number of steps, even though DFT calculations are used solely for external validation and do not enter the optimization loop as a prior. Notably, a pronounced reduction in energy is observed already at the early stages of the optimization, closely mirroring the rapid decrease of the \textit{R}-factor (Fig.~\ref{fig:rf}a). This concurrent decrease stems from substantial structural adjustments during the initial stage (Fig.~\ref{fig:rf}b, c), revealing a notable consistency between LEED-$I(V)$ matching and DFT energy minimization. Specifically, as illustrated in Stages A and B of Fig.~\ref{fig:rf}c, the silver atoms move within the $\textbf{\textit{x}}$-$\textbf{\textit{y}}$ plane toward the vertices (coordinate (0, 0)) and the center (coordinate (0.5, 0.5)) of the supercell. This collective alignment behavior enhances the structural symmetry of the system, thereby further reducing its total energy. This behavior demonstrates that even when the \textit{R}-factor is employed as the sole objective function, the optimization trajectory naturally evolves toward energetically favorable configurations. This synergy indicates that the \textit{R}-factor implicitly encodes rich physical information about the surface structure, effectively guiding the search toward physically realistic solutions without explicit energetic constraints.

However, the evolutionary trends in the structure's total energy and the \textit{R}-factor are not entirely consistent. In the middle to late stages of optimization, a decoupling phenomenon occurs in the parameter optimization. Specifically, careful examination of the energy trajectory (Fig.~\ref{fig:energy}a) reveals that after approximately the 280 sampling steps, while the \textit{R}-factor continues to decline, the DFT total energy shows a slight increase. Concurrently, the VIBROCC parameters exhibit significant fluctuations during the same phase (Fig.~\ref{fig:rf}a). This energy increase reflects a shift in optimization focus from structural parameters to non-structural parameters, rather than a loss of physical plausibility. From a Bayesian perspective, the likelihood function can be expressed as:

\begin{equation}
   p(\mathcal{D}|\Theta) \propto \exp\!\left[-\alpha R(\Theta)\right],
\label{eq:likelihood}
\end{equation}
where $\mathcal{D} = Y_{\mathbf{i}}^{\mathrm{expt}}(E)$ denotes the experimental data and $\alpha$ is a scaling coefficient. We further decompose the parameter space into structural coordinates $\Theta_c$ and non-structural parameters $\Theta_s$ (VIBROCC). If the DFT energy were explicitly incorporated as a prior on the structural coordinates,
\begin{equation}
   p(\Theta_c) \propto \exp\!\left[-\beta E_{\mathrm{DFT}}(\Theta_c)\right],
\label{eq:pri}
\end{equation}
Then the posterior becomes:

\begin{equation}
   p(\Theta|\mathcal{D}) \propto \rm{exp}(-\alpha R(\Theta)-\beta E_{DFT}(\Theta_c))
\label{eq:pri}
\end{equation}
where $\rm E_{DFT}$ is the DFT-calculated energy and the $\beta$ is the scaling coefficient. However, we do not explicitly incorporate energy into the prior, which leads to, during the later stages of optimization $\nabla_{\Theta_c} R(\Theta) \approx 0 $ but $\nabla_{\Theta_s} R(\Theta) \neq 0 $. This behavior is clearly reflected in Fig.~\ref{fig:rf}a, where pronounced fluctuations in VIBROCC appear around the 280th sample. Considering the marginal likelihood:

\begin{equation}
   p(\mathcal{D}|\Theta_s) = \int p(\mathcal{D}|\Theta_c,\Theta_s)d\Theta_c \approx const.\times \rm{exp}(-\alpha R(\Theta_s)
\label{eq:mar}
\end{equation}
At this stage, the optimization is dominated by $\Theta_s$, and BO sampling naturally concentrates along the direction of non-structural parameters. This explains why structures with slightly higher DFT energy can still achieve marginally lower \textit{R}-factors in the later stages of optimization, and also demonstrates the algorithm's ability to adaptively decouple parameters without requiring manual intervention in the optimization process.

Furthermore, the optimized solutions of the LEED-$I(V)$ inverse problem are found to reside within a narrow and nearly energy-degenerate structural basin. As shown in Fig.~\ref{fig:energy}b, the \textit{R}-factors and DFT total energies of all sampled structures fall within this region. It is particularly noteworthy that within an energy span slightly exceeding 0.01~eV and an \textit{R}-factor variation range of approximately 0.04 (see the inset in Fig.~\ref{fig:energy}b), a high density of sampled points is clustered, with configurations exhibiting low \textit{R}-factors (less than 0.08) being especially concentrated. Crucially, the total energies within this region span only a small range, indicating that a set of structurally similar, nearly degenerate configurations can reproduce the experimental spectra with comparable accuracy. To quantify this equivalence, we compared two representative cases: Case A (lowest \textit{R}-factor) and Case B (lowest energy). Their key parameters are listed side-by-side in Table~\ref{tab:comparison}. The total energy of the supercell in Case B is –15.3469 eV, merely 0.6 meV lower than that of Case A (–15.3463 eV). This energy difference is physically negligible and indicates that the two structures are essentially equivalent. More specifically, the corresponding optimized atomic coordinates of the two cases are nearly identical, and the resulting LEED-$I(V)$ spectra are almost indistinguishable (Fig.~\ref{fig:energy}c). Therefore, the optimal structural solution was essentially determined around the 280 to 300 sampling steps (Fig.~\ref{fig:rf}a), with subsequent reductions in the \textit{R}-factor stemming from fine-tuning of the VIBROCC parameters. Based on its slightly superior energetic stability, we adopt the structure of Case B as the final reconstruction model. In general, such a behavior highlights a fundamental characteristic of inverse problems constrained by finite experimental resolution: below the resolution limit, families of nearly degenerate solutions naturally emerge. At the same time, the observed clustering demonstrates the ability of the proposed framework to autonomously identify the narrow confidence basin containing the physically relevant solutions through adaptive TR control, without requiring manual intervention or heuristic constraints.

Based on the analysis above, we further confirmed through a controlled experiment that surface-specific vibrational amplitudes are indispensable for accurate LEED-$I(V)$ structure determination and cannot be approximated by bulk values. To directly validate the independence and influence of surface vibrational parameters, we constructed a comparative Case C (Fig. \ref{fig:energy}c) by fixing the VIBROCC to the bulk value  (0.069~\si {\angstrom}) based on the energy-optimal structure of Case B. The resulting \textit{R}-factor soared to 0.4029—more than five times higher than the optimized values of Cases A and B, far exceeding the acceptable fitting range. Although the LEED-$I(V)$ fit captured the geometric features of the spectra, it exhibited significant intensity deviations compared to the experimental spectra, as shown in Fig.~\ref{fig:energy}c. This marked deterioration demonstrates that structural optimization at the DFT level alone is insufficient for quantitatively reproducing LEED-$I(V)$ spectra. Specifically, standard DFT provides equilibrium structures at 0 K, whereas LEED-$I(V)$ experiments inherently probe surfaces at finite temperatures, where thermal vibrations significantly modulate scattering intensities. This also explains why the DFT-optimized structure differs to some extent from the structure obtained from LEED-$I(V)$ analysis in Ref.~\cite{tianTwoDimensionalMetalPhosphorusNetwork2020}—that is, this difference arises from the presence of atomic thermal vibration parameters. Therefore, reliable reconstruction requires the explicit and simultaneous optimization of both geometric parameters and thermal vibrational properties. 

\begin{table}[htbp]
\centering
\caption{Comparison of structural and energetic parameters obtained from two optimizations.}
\label{tab:comparison}
\begin{tabular*}{\textwidth}{@{\extracolsep\fill}lcc}
\toprule
\textbf{Parameter} & \textbf{Case A} & \textbf{Case B} \\
\midrule
\textit{R}-factor            & 0.0772          & 0.0788          \\
Energy          & -15.3463 (eV)    & -15.3469 (eV)    \\
VIBROCC  
                    & 0.09479 (\si{\angstrom})         & 0.09433 (\si{\angstrom})        \\
\midrule
\multicolumn{3}{l}{\textbf{Optimized fractional coordinates}} \\
\midrule
Atom 1 & (0.00000, 0.020466, 0.799707) & (0.00000, 0.020467, 0.799789) \\
Atom 2 & (0.500000, 0.518898, 0.701105) & (0.500000, 0.518459, 0.701182) \\
Atom 3 & (0.000000, 0.006003, 0.600405) & (0.000000, 0.007095, 0.600462) \\
Atom 4 & (0.500000, 0.494911, 0.500230) & (0.500000, 0.494911, 0.500243) \\
\bottomrule
\end{tabular*}
\end{table}

\subsection{BO boosted autonomous structure determination of Fe\textsubscript{2}O\textsubscript{3}(1$\overline{1}$02)-(1×1) surface}\label{subsec2}

\begin{figure}[H]
\centering
\includegraphics[width=1\textwidth]{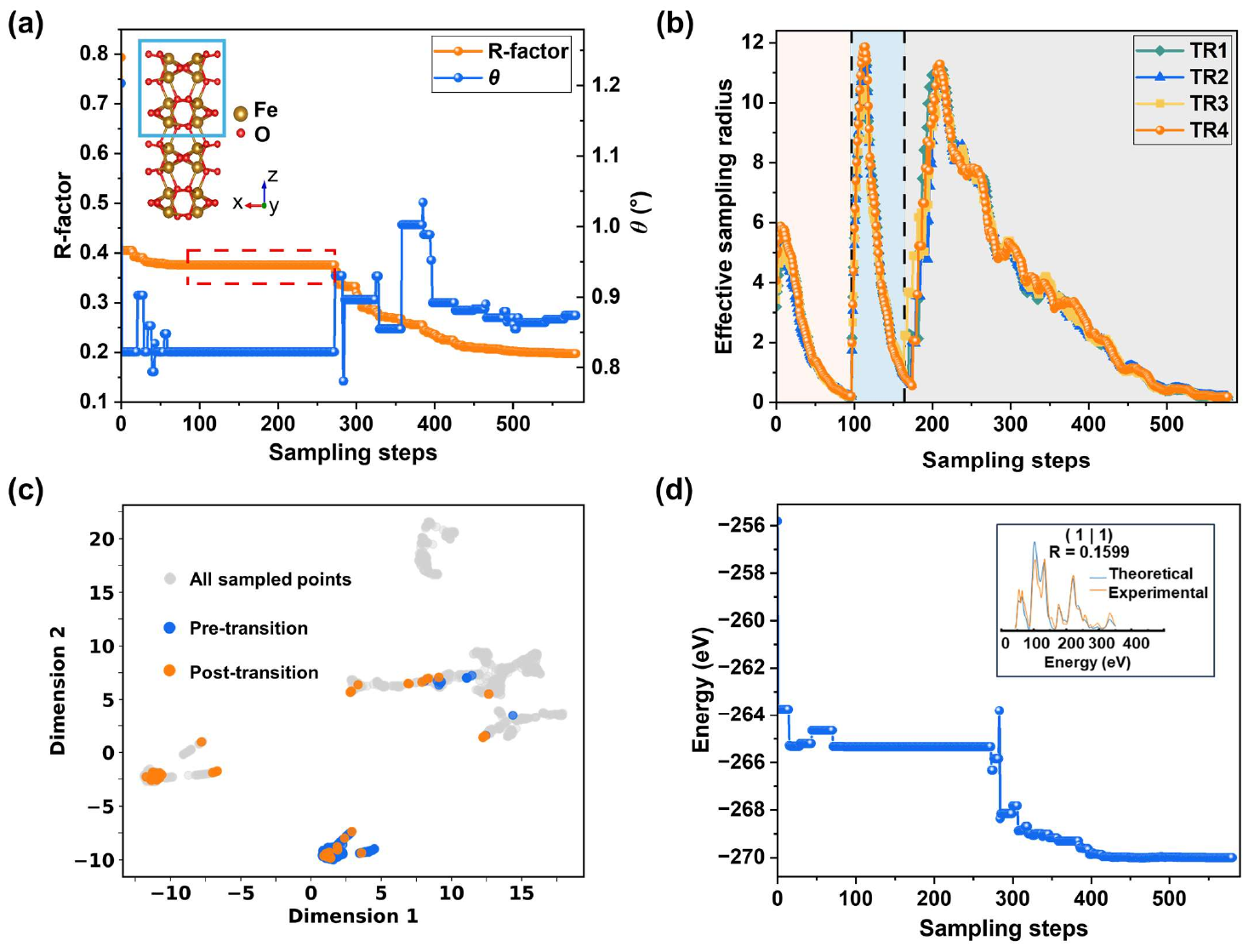}
\caption{\textbf{Convergence and structural evolution during autonomous surface reconstruction of Fe\textsubscript{2}O\textsubscript{3}(1$\overline{1}$02)-(1×1) surface}. (a) Evolution of the \textit{R}-factor and the electron beam incidence angle $\theta$ as a function of sampling steps. The Fe\textsubscript{2}O\textsubscript{3}(1$\overline{1}$02)-(1×1) surface slab is shown to indicate the region where only the atomic coordinates within the blue box are allowed to vary during the optimization. The red dashed box indicates the stage where the code gets trapped in a local optimum. (b) Variation of the effective sampling radius with increasing sampling steps across four TRs. The pink, light blue, and gray colors indicate the three stages of gradually contracting sampling radius. Two black lines indicate the position where the transition in the effective sampling radius occurs. (c) UMAP distribution of the sampled data points. The gray points represent all sampled configurations, while the blue points indicate the samples collected before the transition at around 100 sampling steps, and the orange points represent the 64 samples collected after the transition at around 100 sampling steps. The highlighted subset corresponds to the transition zone between the pink and light blue regions in panel (b), reflecting the adaptive shift of the sampling area during the optimization process. (d) Evolution of the DFT total energy of Fe\textsubscript{2}O\textsubscript{3}(1$\overline{1}$02)-(1×1) along the sampling trajectory, with an inset showing the fitting result of the final $(1 \mid 1)$ beam.}
\label{fig:fe}
\end{figure}

For complex surface LEED-$I(V)$ spectra, our algorithm also demonstrates sufficient robustness. Fig.~\ref{fig:fe} shows the reconstruction result of the Fe\textsubscript{2}O\textsubscript{3}(1$\overline{1}$02)-(1×1) surface. This inverse problem is defined in a high-dimensional parameter space of 53 variables, including atomic displacements (for 27 atoms with fractional coordinates greater than 0.25), thermal vibration factors (VIBROCC$_{\rm Fe}$, VIBROCC$_{\rm O}$, with a variation range of 0.1~\si{\angstrom}), and the electron beam incidence angle $\theta$. Notably, the inclusion of $\theta$ as an optimizable parameter enables the framework to autonomously compensate for experimental misalignment within a range of $0^\circ$–$2^\circ$, a task that is conventionally required to be decoupled from structural refinement. Detailed information on the optimization process is provided in Table S2. This case confirms the scalability and robustness of our proposed physics-informed BO method when dealing with complex surface structures.

Despite the ruggedness of the high-dimensional parameter landscape, the algorithm successfully escapes metastable solutions and converges to a physically meaningful optimum. As indicated by the red dashed box in Fig.~\ref{fig:fe}a, the optimization process encountered a distinct local optimum, with the optimal \textit{R}-factor stagnating at 0.3757 from around 80 to around 280 sampling steps. During this plateau phase, the sampling geometry gradually transitions to a broader search regime. To understand what is happening, we introduce the effective sampling radius to directly characterize this behavior from the query distribution. For a sliding window of size $k$, let $\boldsymbol{W}_t = \{\boldsymbol{x}_{t-k+1}, \ldots, \boldsymbol{x}_t\}$ denote the queried data points and:

\begin{equation}
  \boldsymbol{c}_{t} = \frac{1}{|\boldsymbol{W}_t|} \sum_{\boldsymbol{x}_{t-k+1} \in \boldsymbol{W}_t} \boldsymbol{x}_{t-k+1}
\end{equation}
their local centroid. The effective sampling radius is defined as:

\begin{equation}
  R_{\rm{eff}}(t) = \frac{1}{|\boldsymbol{W}_t}| \sum_{\boldsymbol{x} \in \boldsymbol{W}_t} \| \boldsymbol{x} - \boldsymbol{c}_t \|^2
\end{equation}
which quantifies the realized geometric dispersion of the sampled points. Notably, this metric reflects the actual distribution of queried data points and therefore captures the realized search scale. As shown in Fig.~\ref{fig:fe}b, the effective sampling radius exhibits two transitions between approximately 80 and 280. As marked by the two dashed lines, the effective sampling radius rises sharply at around 100 and 170 sampling steps for all four TRs, indicating a significant expansion of the effective sampling range. Fig.~\ref{fig:fe}c displays the UMAP distribution of all sampling points, as well as the changes in the UMAP distribution of all points before the transition at approximately 100 steps, compared to that of the 64 sampling points after the transition. The post-transition samples span a substantially broader region in the reduced feature space compared to the pre-transition samples, demonstrating a clear shift from a localized to a dispersed sampling regime. This expansion of the sampling radius enables the proposal of new candidate points at greater distances from the current local region. From a geometric perspective, this is equivalent to increasing the accessible volume in parameter space, thereby enhancing the probability of traversing the barriers that separate different local minima. The animation in the supplementary materials clearly shows the variation in the UMAP distribution of sampling points throughout the optimization process, which can help us better understand this adaptive exploration mechanism.

As a result of this adaptive exploration strategy, the algorithm successfully converges to a solution slightly better than that of ViperLEED.calc~\cite{ExamplesViPErLEED0141}: the \textit{R}-factor converged to 0.1977, and the incident angle converged to $0.8740^\circ$, as shown in Fig.~\ref{fig:fe}a. The final atomic positions obtained by the BO method show only minor deviations from those derived by ViperLEED.calc, with maximum fractional coordinate differences of 0.006687, -0.013444, and 0.00025 along the three directions (see Table S3), respectively. All final optimized parameters are summarized in Table S3. Meanwhile, the final fitting quality of LEED-$I(V)$ (for the (1 $\mid$ 1) beam shown in Fig.~\ref{fig:fe}d, and all other beams' results provided in Fig. S5) indicates that despite the high-dimensional optimization, the calculated spectra remain in excellent agreement with the experimental data. The convergence of structural energy further confirms the physical consistency of the proposed algorithm. As illustrated in Fig.~\ref{fig:fe}d, the total energy (calculation details in Table S2) evolves systematically as the number of samples increases, reaching its minimum value simultaneously with the convergence of the \textit{R}-factor. A supplementary animation better illustrates the evolution of the \textit{R}-factor, spectral fitting quality, and structural transformations during the optimization process. Collectively, these results demonstrate that the proposed framework can reliably solve high-dimensional, strongly coupled LEED-$I(V)$ inverse problems within a single, unified workflow that simultaneously optimizes structural, vibrational, and experimental geometric parameters—without sacrificing the spectral fidelity typically achieved only through multi-stage, manually guided procedures.

In principle, the \textit{R}-factor could be further reduced by increasing the number of atoms allowed to relax, {\it e.g.}, by optimizing all atoms with fractional coordinates greater than 0.14 along the $z$-direction. Such an expansion would raise the number of free parameters from 53 to 78, significantly increasing computational cost. Moreover, the increase in parameter space dimensionality would demand more sampling points per iteration to maintain effective exploration. It is important to emphasize, however, that the present framework is not designed to accelerate the underlying forward LEED calculations. Instead, its central contribution lies in transforming the traditional, expert-driven LEED-$I(V)$ analysis workflow—characterized by heuristic parameter tuning and sequential refinement—into a reproducible, fully automated inverse modeling pipeline.

Consistent with standard LEED-$I(V)$ quantification practice, the present study assumes a physically reasonable initial structural model. Relying solely on \textit{R}-factor feedback, the method autonomously explores the coupled parameter space, thereby eliminating reliance on human intuition, manual intervention, or predefined optimization hierarchies. This enables us to systematically and unbiasedly explore high-dimensional parameter spaces, a task that would be extremely tedious and laborious if tackled using traditional multi-step screening strategies. Beyond LEED-$I(V)$, the generality of the framework makes it readily extensible to other (surface-sensitive) characterization techniques governed by physics-based forward models. That is, for characterization techniques where physical simulations can be performed, our BO approach can be applied to tackle the inverse quantization problem, given parameter spaces of reasonable dimensions. Moreover, our approach allows integrating diverse relevant experimental characterization techniques, {\it e.g.}, combining LEED-$I(V)$ with SXRD or STM. This opens the door to multimodal surface structure quantification, where consistency among multiple physical probes yields more robust and physically realistic structural models.

Importantly, the comparison with DFT-calculated structural energies in this study further validates the physical plausibility of the proposed structure, even though such energetics has not been explicitly included as a prior in the algorithm. This suggests a promising strategy to streamline quantification processes by incorporating physical constraints, {\it i.e.}, pre-filtering energetically favorable candidate structures. This can be an essential step for complex systems like twisted moiré superlattices~\cite{yinEpitaxialGrowthMono2025,qiuFormationLargeAreaTwisted2023,liDirectCVDGrowth2025}. 
We foresee such an extension being facilitated using machine-learned interatomic potentials (MLIPs), which not only overcome the bottleneck of DFT for large systems with tens of thousands of atoms~\cite{siddiquiUnderstandingDomainReconstruction2025,sauerDispersioncorrectedMachineLearning2025,wangAtomicClusterExpansion2025} but also form a powerful synergy for surface characterizations. 
For instance, by simulating the relaxation behavior of two-dimensional moiré systems using machine-learned potentials, height modulations, strain distributions, and electronic phase transitions observed in STM~\cite{shaiduTransferableDispersionawareMachine2025a}, Raman~\cite{deyMemorizationStrainInducedMoire2025}, and other experiments can be theoretically reproduced. 
Therefore, the synergy between MLIPs and surface characterization not only significantly enhances the reliability of atomic-level structural models but also accelerates the closed-loop process from experimental observations to designable properties~\cite{deyMemorizationStrainInducedMoire2025, lunaMachineLearningInteratomicPotential2025, magorrianStrongAtomicReconstruction2025a}. 
Consequently, a multi-fidelity BO framework that integrates MLIPs with characterization data holds great potential for achieving more efficient and robust structural reconstruction, which is saved for future study.

\section{Conclusion}\label{sec4}

In summary, we have reformulated quantitative LEED-$I(V)$ structure determination as a physics-constrained inverse problem and achieved an autonomous solution by implementing a BO framework that incorporates forward models based on multiple scattering. Such a physics-informed automated LEED-$I(V)$ quantification framework preserves the rigor of scattering simulations while enabling efficient exploration of high-dimensional coupled parameter spaces. Unlike traditional expert-driven workflows, our approach relies exclusively on \textit{R}-factor feedback to guide the optimization, eliminating the need for human intuition, manual parameter tuning, or predefined refinement hierarchies. We validated the framework across two benchmark cases of differing complexity: the well-defined Ag(100)-(1×1) surface and the more complex Fe\textsubscript{2}O\textsubscript{3}(1$\overline{1}$02)-(1×1) surface, for which it robustly reconstructs three-dimensional surface structures within a single, unified workflow. Independent validation through first-principles-derived energetics analysis confirms that the \textit{R}-factor–driven search naturally converges toward energetically favorable configurations, underscoring the physical consistency of the inferred solutions. Beyond LEED-$I(V)$, the proposed framework can be easily extended to other characterization techniques, and the demonstrated methodology establishes BO as a general-purpose, physics-preserving engine for solving the inverse problems in experimental characterization techniques. This marks a transformation from a conventional expertise-driven process to an autonomous, physics-driven quantification, paving the way for reproducible, scalable, and fully automated material characterizations.

\section*{Acknowledgements}

The authors gratefully acknowledge the computing time provided on the high-performance computer Lichtenberg at the NHR Centers NHR4CES at TU Darmstadt. Xiankang Tang thanks the financial support from the China Scholarship Council.

\section*{Data and Code Availability Statements}

The codes are available at the repository on GitHub (\url{https://github.com/XiankangTang/BO_LEED})

\bibliography{LEED}

\end{document}